# Hate multiverse spreads malicious COVID-19 content online beyond individual platform control

N. Velásquez[1,2,3], R. Leahy[1,2,3], N. Johnson Restrepo[1,2,3], Y. Lupu[4,3], R. Sear[2,5], N. Gabriel[2], O. Jha[2], B. Goldberg[6], N.F. Johnson[1,2]
[1]Institute for Data, Democracy and Politics, George Washington University, Washington D.C. 20052
[2]Physics Department, George Washington University, Washington D.C. 20052
[3]ClustrX LLC, Washington D.C.
[4]Department of Political Science, George Washington University, Washington D.C. 20052
[5]Department of Computer Science, George Washington University, Washington D.C. 20052
[6]Google LLC, New York City, NY 10011

**We show that malicious COVID-19 content, including hate speech, disinformation, and misinformation, exploits the multiverse of online hate to spread quickly beyond the control of any individual social media platform. Machine learning topic analysis shows quantitatively how online hate communities are weaponizing COVID-19, with topics evolving rapidly and content becoming increasingly coherent. Our mathematical analysis provides a generalized form of the public health $R_0$ predicting the tipping point for multiverse-wide viral spreading, which suggests new policy options to mitigate the global spread of malicious COVID-19 content without relying on future coordination between all online platforms.**

Controlling the spread of COVID-19 misinformation and its weaponization against certain demographics (e.g. anti-Asian) -- in particular, by the online hate community of neo-Nazis and other extremists -- is now an urgent problem [1-6]. In addition to undermining public health policies, malicious COVID-19 narratives are already translating into offline violence [2,3]. Making matters worse, each social media platform is effectively its own *universe*, i.e. a commercially independent entity subject to particular legal jurisdictions, and hence can at best only control content in its universe [1,4]. Moreover, there is now a proliferation of other, far less moderated platforms thanks to open-source software enabling decentralized setups across locations.

Winning the war against such malicious matter will require an understanding of the entire online battlefield *and* new policing approaches that do not rely on global collaboration between social media platforms. Here we offer such a combined solution. Specifically, Figs. 1 and 2 show how COVID-19 malicious content is exploiting the existing online hate network to spread quickly between platforms and hence beyond the control of any single platform (Fig. 1A,B). Methods and Supplementary Information (SI) give details and examples of this material. Links between distinct platforms (i.e. universes) act like wormholes to create a huge, decentralized multiverse that connects hate communities (nodes with black circles, Fig. 1B) to the mainstream (nodes without black circles, Fig. 1B). Figure 1B involves ~10,000,000 users across languages and continents who have formed themselves into ~6,000 inter-linked public clusters, i.e. online communities such as a Facebook page, VKontakte group, or Telegram channel, each represented as a node in Figs. 1,2. These new insights inform the policy prescriptions offered in Fig. 3.

Our methodology [7, 8] focuses on the mesoscopic scale of online clusters (i.e. node in Fig. 1) where each cluster is an interest-based online community (e.g. VKontakte group). It is known that such clusters are where people develop, and coordinate around, narratives [9] -- in contrast to platforms like Twitter that have no pre-built community tool and are instead designed for broadcasting short messages [7-9]. Each link is an online hyperlink that appears at the level of the entire cluster (e.g. Fig. 2A). Including links between clusters across different platforms (see Methods and SI) then enables us to map the broader, global online ecology at the entire system level.

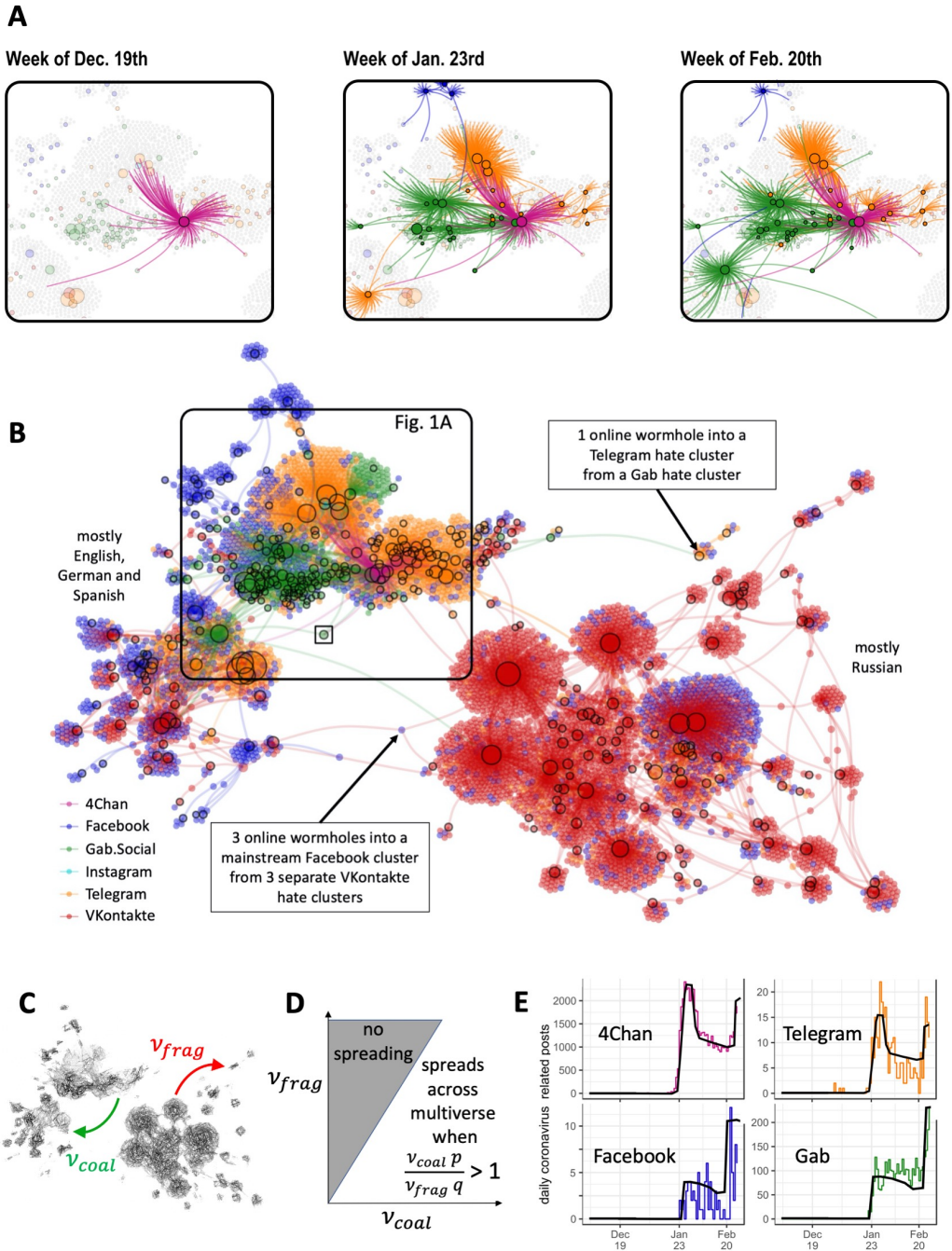

Figure 1: Spreading across Online Hate Multiverse. A: Time evolution of birth and spread of malicious COVID-19 content within and across different social media platforms (i.e. universes) for small section from B.
B: Online hate multiverse comprises separate social media platforms (i.e. universes) that interconnect over time via dynamical links (i.e. wormholes) created by hyperlinks from clusters on one platform into clusters on another (e.g. Fig. 2A). Links shown are from hate clusters (i.e. online communities with hateful content, shown as a node with a black ring) to all other clusters including mainstream ones (e.g. football fan club) which it can then influence. Link color denotes the platform hosting the hate cluster from which link originates. Plot aggregates activity from June 1st 2019 to March 23rd, 2020. The observed layout is spontaneous (i.e. not built-in, see Methods). Small black square is a Gab cluster analyzed in Fig. 2B. C: Model features dynamical links connecting and disconnecting clusters of clusters (i.e. coalescence with probability $\nu_{coal}$, fragmentation with probability $\nu_{frag}$). D: Phase diagram shows generalization of public health $R_0$ that predicts tipping point for online spreading. E: Output from model in C,D compared to empirical data from A (see Methods and SI for details).

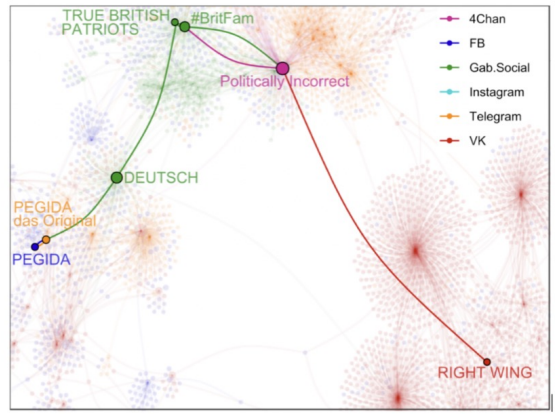

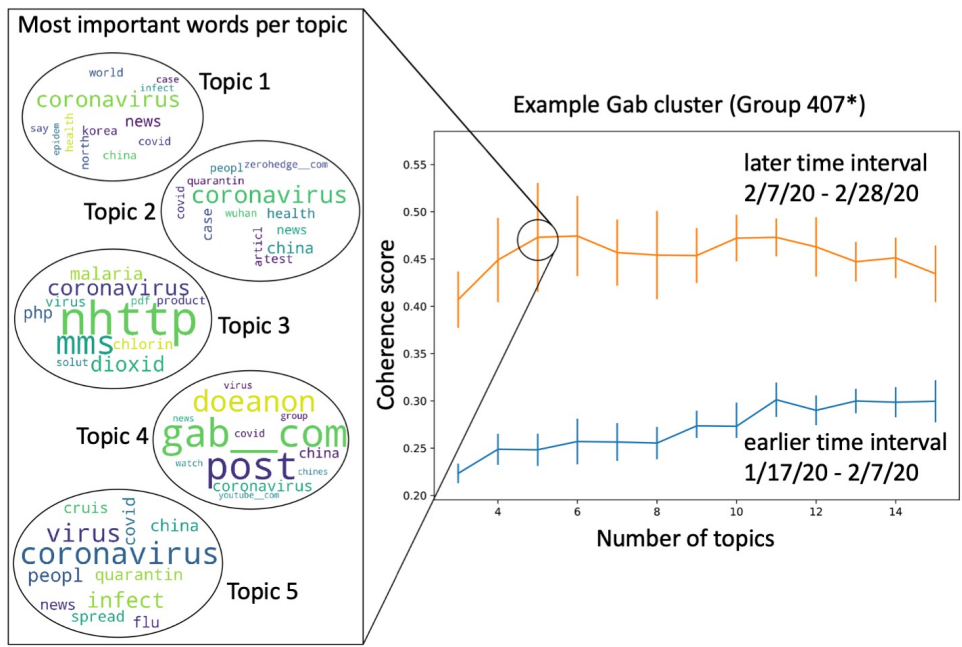

**Figure 2: Multiverse Pathways and Content. A:** Example pathway for a piece of malicious matter. **B:** Example output from our machine learning topic analysis of content [10] within a single example Gab cluster (see small black circle in Fig. 1B). Even though COVID-19 topic only arose in December 2019, it quickly evolved from featuring a large number of topics with a relatively low average coherence score, a measure of semantic similarity, to a smaller number of topics with higher average coherence scores more focused around COVID-19. Reflecting this, we note that prior to COVID-19, topics featured words like f*** and n*****, while the conversation around COVID-19 is more focused and less like a stereotypical hate-speech rant. SI shows explicit examples of this content.

The first general implication of our findings is that policies to curb COVID-19 and related malicious matter need to account for the decentralized, interconnected nature of this multiverse (Fig. 1). Links connecting nodes from different universes (i.e., different social media platforms) provide a gateway that can pass malicious content (and supporters) from a cluster on one platform to a cluster on another platform that may be very distant geographically, linguistically, and culturally, e.g. from Facebook to VKontakte. Figure 2A shows that consecutive use of these links allows malicious matter to find short pathways that cross the entire multiverse, just as short planks of wood can be used to bridge adjacent rocks and cross a wide river. Moreover since malicious matter frequently carries quotes and imagery from different moments in a cluster's timeline, these inter-platform links not only interconnect information from disparate points in space, but also time -- like a wormhole.

A second implication comes from our machine-learning topic analysis using Latent Dirichlet Allocation to identify topics discussed in the online hate multiverse, and then calculating a coherence score for different topics [10] (see Fig. 2B and SI). This shows that clusters in the global online hate community are coalescing around COVID-19, with topic flavors evolving rapidly and their coherence scores increasing. Examples of weaponized content (see SI) reveal evolving narratives such as blaming Jews and immigrants for inventing and spreading the virus, and instances of neo-Nazis planning attacks on emergency responders to the health crisis. While these topics have evolved over , the underlying structure in Fig. 1B remains rather robust which suggests that our implications should also hold in the future.

A third implication is that malicious activity can *appear* isolated and largely eradicated on a given platform, when in reality it has moved through a wormhole to another universe. There, malicious content can thrive beyond that platform's control, be further honed, and later reintroduced into the original platform using a wormhole in the reverse direction. Moderators reviewing only blue clusters in Fig. 1B might conclude that they had largely rid that platform of hate and disconnected hateful pages from one another, when in fact these same clusters remain connected via other universes. Because the number of independent online universes (i.e. social media platforms) is growing, this multiverse will continue to grow and will likely be deeply interconnected via new wormhole links.

Implication 4 is that this multiverse acts like a global funnel that can suck individuals from a mainstream cluster on a platform that invests significant resources in moderation, into less moderated platforms like 4Chan or Telegram, simply by following the links offered to them. Critically, an innocent user of mainstream social media communities, including a child connecting with other online game players or a parent seeking information about COVID-19, is at most a few links away from intensely hateful content. In this way, the rise of fear and misinformation around COVID-19 has allowed promoters of malicious matter and hate to engage with mainstream audiences around a common topic of interest, and potentially push them toward hateful views.

Implication 5 is that it is highly unlikely that the multiverse in Fig. 1B is, or could be, coordinated by a single state actor, given its vast decentralized nature. We have checked for signals of Russian-sponsored campaigns. Since many hate clusters organize around the topics of minorities and refugees, we expected to find frequent links to Russian media, but instead only found a small portion of clusters linking to Kremlin-affiliated domains. These links accounted for <0.5% of all posts shared. This is also consistent with the notion that the extended nature of exchanges in a cluster (i.e. online community) enables it to collectively weed out unwanted trolls and bot-like members.

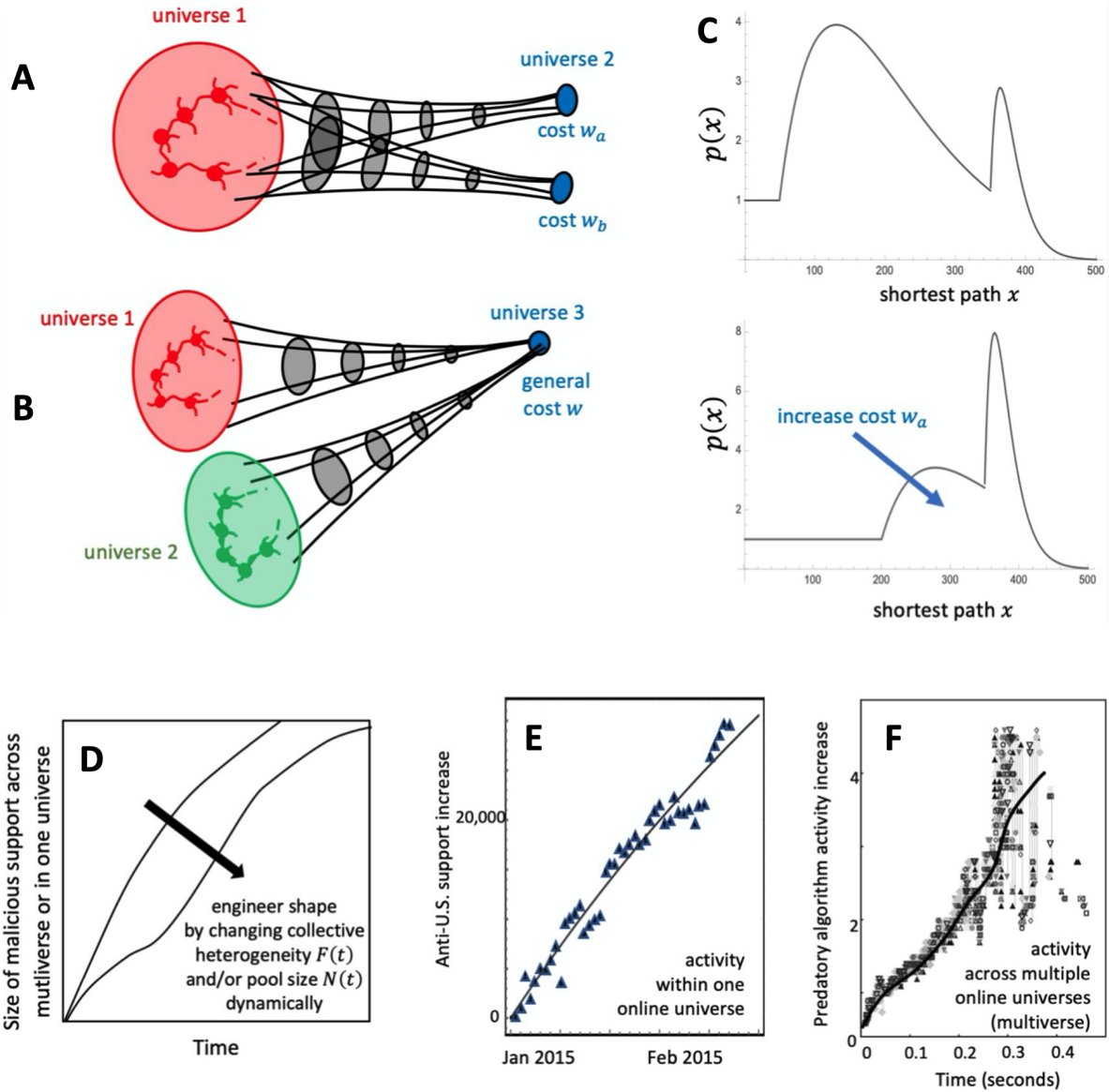

**Figure 3: Wormhole Engineering to Mitigate Malicious Content Spreading. A-B: Typical motifs** within the full multiverse in Fig. 1B.
**C:** Mathematical prediction for motif A, showing that the distribution of shortest paths (top panel, shown unnormalized) for transporting malicious matter across a platform (i.e. universe 1) can be shifted to larger values (bottom panel) which will then delay spreading and will increase the chance that the malicious matter is detected and removed [11,12]. This is achieved by manipulating the risk that the hate material gets detected when passing via the other platform: this risk represents a cost for the hate community in universe 1 when using the blue node(s). Same mathematics applies irrespective of whether each blue node is a single cluster or an entire platform, and applies when both blue clusters are in the same platform or are in different platforms. See SI for case B.
**D-F:** Mathematical prediction that **the total online support for malicious matter can be manipulated by varying the online pool size of potential supporters** $N(t)$ **and/or their heterogeneity** $F(t)$. **E:** Empirical outbreak of anti-U.S. hate across a single platform (VKontakte) produces similar shape to upper curve in D. **F:** Empirical outbreak for the proxy system of predatory 'buy' algorithms across multiple electronic platforms [13] produces similar shape to lower curve in D. (See SI for details).

Implication 6 addresses the key issue that coordinated moderation between all platforms -- while highly desirable -- may not be possible. To bypass this, our mathematical predictions suggest that bilateral wormhole engineering could be used by platforms to artificially lengthen the pathways that malicious matter needs to take between clusters, hence increasing the chances of its detection by regulators and also delaying the spread of time-sensitive material such as hate manifestos and real-time streaming of attacks (see SI for details). This involves the following repeated process: first, pairs of platforms use Fig. 1B to estimate the likely numbers of wormholes between them. Then without having to exchange any sensitive data, each can then use our mathematical formulae (see SI) to engineer the correct cost *w* for hate-spreaders who are exploiting their platform as a pathway, e.g. they focus available moderator time to achieve a particular detection rate for hate material passing through their platform and hence create an effective cost *w* for these hate-spreaders in terms of detection and removal. While Figs. 3A,B show common situations that arise in Fig. 1B, more complex combinations can be described using similar calculations (see SI) in order to predict how the path lengths for hate material can be artificially extended in a similar way to Fig. 3C.

With or without such wormhole engineering, the *dynamics* of how malicious matter passes across the multiverse in Fig. 1B is highly complex since there are wormholes (and hence pathways) opening and closing (or getting restricted) in real time, and hence subsets of clusters effectively coalescing or fragmenting as in Fig. 1C. Despite this complexity, we can provide the condition that needs to be met to *prevent* multiverse-wide spreading of a particular piece of malicious matter (see SI for derivation [14]). This no-spreading condition is $R_0 = (v_{coal}\, p)/(v_{frag}\, q) < 1$ where $v_{coal}$ ($v_{frag}$) is the average rate at which subsets of clusters coalesce (fragment) within and across platforms; *p* is the average rate at which a single cluster shares material with another cluster; *q* is the average rate at which a single cluster becomes inactive. These parameters can be estimated empirically or from simulations. Conversely the condition for system-wide spreading, which can be used to guide dissemination of counter-messaging, is $R_0 > 1$. While *p* and *q* are properties related to a single average cluster and likely harder to manipulate, platform engineers can use the tools at their disposal to try to change $v_{coal}$ and $v_{frag}$ and hence engineer the value of $R_0$. If no such wormhole engineering can be arranged, our predictions (see SI) show that an alternative though far more challenging way of reducing the spread of hate material is by manipulating either (1) the size *N* of its online potential supporters (e.g. by placing a cap on the size of clusters) and/or (2) their heterogeneity *F* (e.g. by introducing other content that effectively dilutes a cluster's focus). Figure 3D shows examples of the resulting time-evolution of the online support, given by $N\left(1 - W\left(\left[\frac{-2Ft}{N}\right] exp\left[\frac{-2Ft}{N}\right]\right) / \left[\frac{-2Ft}{N}\right]\right)$ where the resulting delayed onset time for the rise in support is $t_{onset} = \frac{N}{2F}$ and where *W* is the Lambert function [15]. Figures 3E and F show related empirical findings which are remarkably similar to Fig. 3D. Figure 3F is a proxy system [13] in which ultrafast predatory algorithms began operating across electronic platforms to attack a financial market order book in subsecond time (see Ref. 13). Hence Fig. 3F also serves to show what might happen in the future if the hate multiverse in Fig. 1B were to become populated by such predatory algorithms whose purpose is now to quickly spread malicious matter. Worryingly, Fig. 3F shows that this could result in a multiverse-wide rise in malicious matter on an ultrafast timescale that lies beyond human reaction times [13].

This analysis of course requires follow-up work. Our mathematical formulae are, like any model, imperfect approximations. However, we have checked that they agree with large-scale numerical simulations [11-15] and follow similar thinking to other key models in the literature [16-18]. Going forward, other forms of malicious matter and messaging platforms need to be included. However, our initial analysis suggests similar findings for any platforms that allow communities (i.e., clusters) to form. We should also further our analysis of the time-evolution of cluster content using the machine-learning Local Dirichlet Allocation approach and other methods. We could also define links differently, e.g. numbers of members that clusters have in common. However, such information is not publicly available for some platforms, e.g. Facebook. Moreover, our prior study of a Facebook-like platform where such information was available, showed low/high numbers of common members reflects the absence/existence of a cluster-level link, hence these quantities indeed behave similarly to each other. People can be members of multiple clusters. However our prior analyses suggest only a small percentage are active members of multiple clusters. In terms of how people react to intervention, it is known that some may avoid opposing views [19] while for others it may harden beliefs [20]. However, what will actually happen in practice remains an empirical question.

**Methods**
Our methodology for identifying clusters and links builds on Refs. 7, 8 and includes links between clusters across multiple platforms. More details are provided in the SI. The clusters are interest-based online communities (e.g., VKontakte group). We include mainstream platforms like Facebook, VKontakte, and Instagram, that have and enforce policies against hate speech, as well as fringe platforms with minimal content policies like Gab, Telegram, and 4Chan. While the method can be replicated for any topic, Fig. 1B focuses on hate and hate-speech defined as either (a) content that would fall under the provisions of the United States' Code regarding hate crimes or hate speech according to Department of Justice's guidelines, or (b) content that supports or promotes Fascist ideologies or regime types (i.e. extreme nationalism and/or racial identitarianism). On-line communities promoting hate have become prevalent globally and are being linked to many recent violent real-world attacks, including the 2019 Christchurch shootings. We observe many different forms of hate adopting similar cross-platform tricks. Our method focuses on clusters at the mesoscale and posts at the microscale, thus the only data from individuals it captures is the content of their posts, just as information about a specific molecule of water is not needed to describe the bubbles (i.e., clusters of correlated molecules) that form in boiling water. We define a cluster (e.g. Facebook fan page, VKontakte club) as a hate cluster if at least 2 out of 20 of its most recent posts at the time of classification align with the above definition of hate. Whether a particular cluster is strictly a hate philosophy, or simply shows material with tendencies toward hate, does not alter our main findings. Links between clusters are hyperlinks (see for example, Fig. 2A). Our network analysis for Fig. 1B starts from a given hate cluster A and captures any cluster B to which hate cluster A has shared an explicit cluster-level link. We developed software to perform this process automatically and, upon cross-checking the findings with our manual list, were able to obtain approximately 90 percent consistency between manual and automated versions. Figure 2A shows an example of clusters and wormholes between them, from our analysis. Figure 1E shows typical output from our model (Fig. 1C,D) with $\nu_{coal} = 0.95$, $\nu_{frag}=0.05$, $p = 0.05$ for all four panels. For 4Chan and Telegram $q = 0.01$; for Gab and Facebook $q = 0.005$. All four fits use these same two model outputs suitably scaled. Output is smoothed over timepoints like the empirical data which is collected daily. Better fits can be obtained by optimizing parameter choice but our purpose here is just to show that typical output from our model captures the observed features of the empirical spreading. All but one node in Fig. 1B is plotted using the ForceAtlas2 algorithm, which simulates a physical system where nodes (clusters) repel each other while links act as springs, and nodes that are connected through a link attract each other. Hence nodes (clusters) closer to each other have more highly interconnected local environments while those farther apart do not. The exception to this Force Atlas2 layout in Fig. 1B is Gab group 407* ("Chinese Coronavirus", https://gab.com/groups/407*), see small black square in Fig. 1B) which was manually placed in a less crowded area to facilitate its visibility. This particular cluster was created in early 2020 with a focus on discussing the COVID19 pandemic -- however, it immediately mixed hate with fake news and science, as well as conspiratorial content.